\documentclass[prc,twocolumn,tightenlines,showpacs,floatfix]{revtex4}
\usepackage{bm}

\usepackage{graphicx,epsfig}

\usepackage{dcolumn}     

\def\beq{\begin{equation}}
\def\eeq{\end{equation}}
\def\beqy{\begin{eqnarray}}
\def\eeqy{\end{eqnarray}}

\begin{document}

{

\title{Charge-symmetry breaking forces and isospin mixing in $\bm{^8{\rm Be}}$}

\author{R. B. Wiringa$^1$}
\email{wiringa@anl.gov}
\author{S. Pastore$^2$}
\email{PASTORES@mailbox.sc.edu}
\author{\mbox{Steven C. Pieper$^1$}}
\email{spieper@anl.gov}
\author{Gerald A. Miller$^3$}
\email{miller@phys.washington.edu}
\affiliation{
$^1$Physics Division, Argonne National Laboratory, Argonne, Illinois 60439 \\
$^2$Department of Physics and Astronomy, University of South Carolina, Columbia,
South Carolina 29208 \\
$^3$Department of Physics, University of Washington, Seattle, Washington 98195
}

\date{\today}

\begin{abstract}
We report Green's function Monte Carlo calculations of isospin-mixing (IM)
matrix elements for the $2^+$, $1^+$, and $3^+$ $T$=0,1 pairs of states at 
16-19 MeV excitation in $^8$Be.
The realistic Argonne $v_{18}$ (AV18) two-nucleon and Illinois-7 
three-nucleon potentials are used to generate the nuclear wave functions.
Contributions from the full electromagnetic interaction and strong 
class III charge-symmetry-breaking (CSB) components of the AV18
potential are evaluated.
We also examine two theoretically more complete CSB potentials based on 
rho-omega mixing, tuned to give the same neutron-neutron scattering length
as AV18.
The contribution of these different CSB potentials to the $^3$H-$^3$He,
$^7$Li-$^7$Be, and $^8$Li-$^8$B isovector energy differences is evaluated 
and reasonable agreement with experiment is obtained.
Finally, for the $^8$Be IM calculation we add the small class IV CSB terms 
coming from one-photon, one-pion, and one-rho exchange, as well as rho-omega 
mixing.
The expectation values of the three CSB models vary by up to 20\% in the
isovector energy differences, but only by 10\% or less in the IM matrix element.
The total matrix element gives 85--90\% of the experimental IM value 
of -145 keV for the $2^+$ doublet, with about two thirds coming from the
Coulomb interaction.
We also report the IM matrix element to the first $2^+$ state at 3 MeV
excitation, which is the final state for various tests of the Standard 
Model for $\beta$-decay.
\end{abstract}


\maketitle

}

\section {Introduction}
\label{sec:intro}

\begin{figure}[!hbt]
\epsfig{file=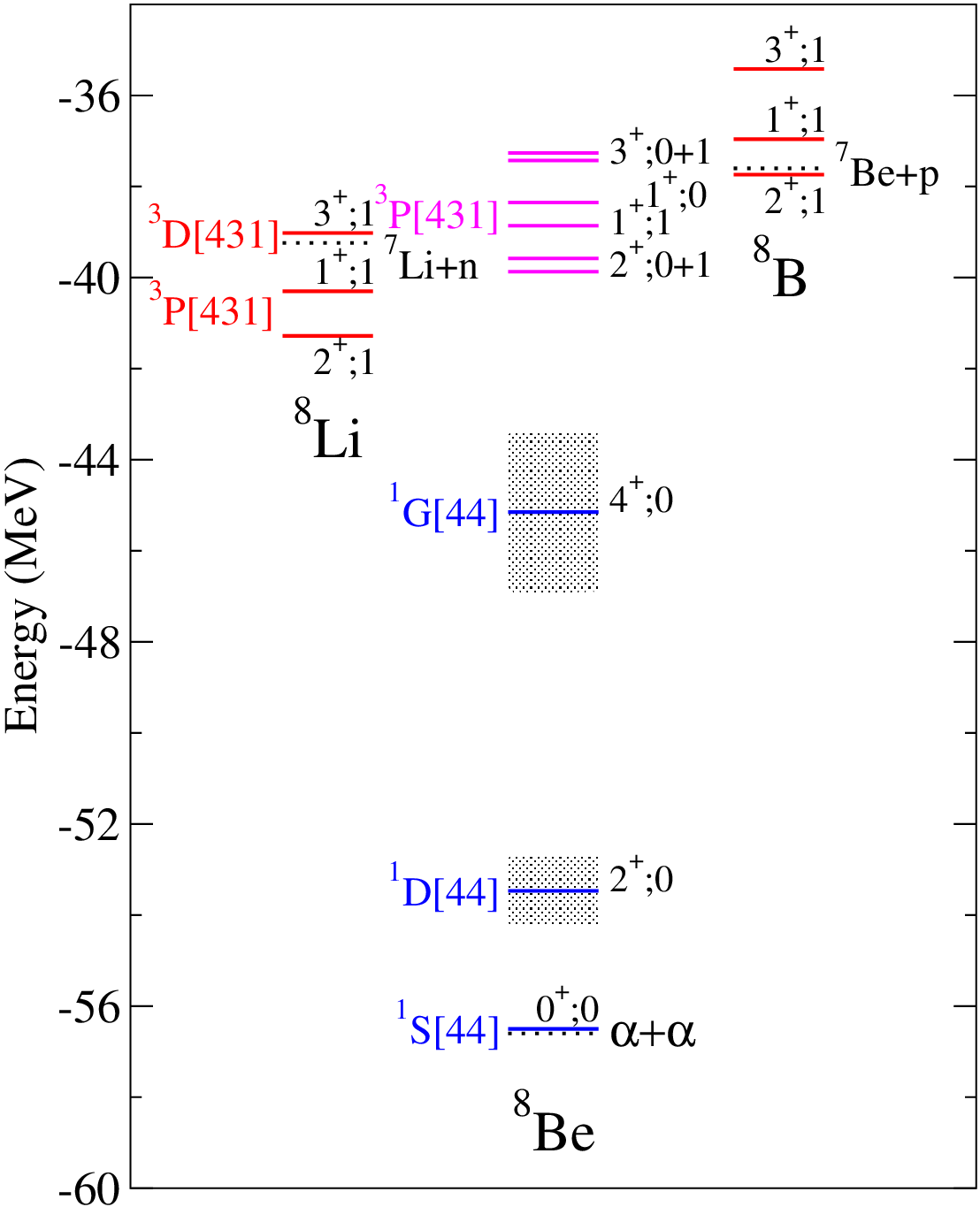,width=8.0cm}
\caption{(Color online) Experimental spectrum of $A$=8 nuclei:
blue lines are $T$=0 states, red lines are $T$=1 states, and
magenta lines are mixed $T$=0+1 states. Dotted lines indicate the
thresholds for breakup and shaded areas denote the large
widths of the $^8$Be rotational states.}
\label{fig:A8}
\end{figure}

The $^8$Be nucleus has a unique excitation spectrum among the light nuclei, 
exhibiting a low-lying rotational band topped by a set of three
isospin-mixed doublets. 
The experimental spectrum~\cite{exp8-10} for low-lying states in $^8$Be and
its isobaric neighbors $^8$Li and $^8$B is shown in Fig.~\ref{fig:A8}.
The structure of these nuclei is well understood on the basis of the allowed
spatial symmetries and spin-isospin combinations.
Realistic nucleon-nucleon forces are strongly attractive in 
relative $S$ waves, hence the most spatially symmetric states will be the most
tightly bound because they maximize the number of $S$-wave pairs~\cite{W06}.
For $^8$Be, the most symmetric states are total isospin $T$=0
with the Young diagram spatial symmetry [44].
In $LS$ coupling the allowed combinations within the $p$-shell are the 
$^{2S+1}L_J$ combinations $^1S_0$, $^1D_2$, and $^1G_4$.
These are the dominant pieces of the $J^\pi$=$0^+$ ground-state and the
first $2^+$ and $4^+$ excitations, respectively, as shown in Fig.~\ref{fig:A8}.
The ground state is unstable against breakup into two $\alpha$ particles
by 0.1 MeV, but is a very narrow (6 eV) resonance, while the two
excited states, which have the structure of two $\alpha$ particles rotating 
about each other, have much larger widths of about 1.5 and 3.5 MeV,
respectively.

The next highest spatial symmetry states are [431] in character
and come in both total isospin $T$=0 and 1 combinations.
The $T$=0 states are the spin triplets $^3P_{0,1,2}$, $^3D_{1,2,3}$, 
and $^3F_{2,3,4}$, while the $T$=1 states come both as these spin 
triplets and as spin singlets $^1P_1$, $^1D_2$, and $^1F_3$.
When diagonalized with a realistic Hamiltonian containing nucleon-nucleon
($N\!N$) and three-nucleon ($3N$) potentials in a
microscopic quantum Monte Carlo (QMC) calculation, the first
three [431] $T$=0 states are ordered $2^+$, $1^+$, and $3^+$, with about 1 MeV
separation, and their dominant components are $^3P_2$, $^3P_1$, and $^3D_3$, 
respectively~\cite{WPCP00}.
The first three [431] $T$=1 states have the same ordering, and about
the same spacing, with the only difference being that there is a moderate 
amount of $^1P_1$ mixed into the $1^+$ state.
These $T$=1 states are the isobaric analogs seen in $^8$Li and $^8$B, giving 
their $2^+$ ground states and low-lying $1^+$ and $3^+$ excited states.
The number of $S$-wave $N\!N$ pairs in the [431] symmetry states is the same
in both $T$=0 and 1 combinations, so it is reasonable to expect that these 
states could appear very close to each other in the spectrum of $^8$Be.
Experimentally there are two $2^+$ states at 16.626 and 16.922 MeV excitation,
two $1^+$ states at 17.64 and 18.15 MeV, and two $3^+$ states at
19.07 and 19.235 MeV, and there is strong experimental evidence for
these states being isospin-mixed.

An early detailed analysis of this mixing was given by Barker in the
course of making intermediate coupling shell-model calculations for
light nuclei~\cite{Barker66}.
The eigenfunctions in his study exhibit the same dominant spatial 
symmetry components found in the later QMC calculations.
A clear experimental signal for isospin mixing of the $2^+$ states
is that both decay by 2$\alpha$ emission, which is the only particle-decay
channel that is energetically allowed, and which is available only
through a $T$=0 component in the initial state.
Following Barker, the eigenfunctions $\Psi_a$, $\Psi_b$ of the observed
states may be written as linear combinations of the $T$=0 and 1 wave functions:
\begin{eqnarray}
  \Psi_a & = & \alpha_J \Psi_0 + \beta_J \Psi_1 \ , \nonumber \\
  \Psi_b & = & \beta_J  \Psi_0 - \alpha_J\Psi_1 \ ,
\end{eqnarray}
with $\alpha_J^2 + \beta_J^2 = 1$. 
The mixing parameters are related to the ratio of $\alpha$-decay widths:
\begin{equation}
  \frac{\Gamma_a}{\Gamma_b} = \frac{\alpha_J^2}{\beta_J^2} \ .
\end{equation}
The current experimental values for the widths are $\Gamma_a = 108.1(5)$ keV
and $\Gamma_b = 74.0(4)$ keV for the 16.626 and 16.922 MeV states,
respectively~\cite{exp8-10}.
This implies $\alpha_2 = 0.7705(15)$ and $\beta_2 = 0.6375(19)$.

The eigenenergies $E_a$, $E_b$ (with $E_a<E_b$) are given by
\begin{equation}
  E_{a,b} = \frac{H_{00}+H_{11}}{2}
\pm \sqrt{ \left(\frac{H_{00}-H_{11}}{2}\right)^2 + (H_{01})^2 }
\end{equation}
where $H_{00}$ is the diagonal energy expectation in the pure $T$=0 state,
$H_{11}$ is the expectation value in the $T$=1 state, and 
$H_{01}$ is the off-diagonal isospin-mixing (IM) matrix element that connects
$T$=0 and 1.
The experimental eigenvalues and eigenenergies, imply that
these matrix elements are $H_{00} = 16.746(2)$ MeV, 
$H_{11} = 16.802(2)$ MeV, and $H_{01} = -145(3)$ keV.
These values are very close to those deduced originally by Barker, and
the values of the mixing parameters for the $2^+$ states are supported 
by a variety of other experimental data.

The analysis for the $1^+$ and $3^+$ doublets is somewhat less direct
because multiple decay channels are available.
For the $1^+$ doublet, Barker used the ratio of $M1$ $\gamma$ transitions
from the 17.64 MeV state to the 16.626 and 16.922 MeV $2^+$ states,
which at the time were in the ratio 1:0.07, to deduce
mixing parameters of $\alpha_1 = 0.24$ and $\beta_1 = 0.97$.
More recent experiments and analyses by Oothoudt and Garvey~\cite{OG77}
produce slightly less mixing, with $\alpha_1 = 0.21(3)$ corresponding
to $H_{01} = -103(14)$ keV.
For the $3^+$ doublet, Barker examined the ratio of neutron to proton
widths for the 19.235 MeV state, and deduced mixing parameters
$\alpha_3 = 0.41$, $\beta_3 = 0.91$, and $H_{01} = -63$ keV.
However, according to Oothoudt and Garvey, the data is consistent with
$0.31 < \alpha_3 < 0.52$, corresponding to an IM matrix element ranging 
from $-47$ to $-71$ keV, and we use this as the empirical value.
But Oothoudt and Garvey also find, on the basis of newer experimental data, 
an even broader range of possible mixing, so the experimental situation 
for the $3^+$ doublet is quite unclear.

The energies of the isospin-unmixed states inferred by using these IM
parameters are given in Table~\ref{tab:energy},
along with the experimental energies and the GFMC energies for the
AV18+IL7 Hamiltonian discussed below.

\begin{table}[t]
\caption{GFMC ground state energy and excitations in MeV for the AV18+IL7
Hamiltonian compared to the empirical energies of the isospin-unmixed 
states and the experimental (isospin-mixed) energies of the $^8$Be spectrum;
also given are the GFMC point proton (= neutron) radii in fm.
Theoretical or experimental errors $\ge 1$ in the last digit are shown in 
parentheses.}
\begin{ruledtabular}
\begin{tabular}{l d d d d}
$J^\pi;T$ & \multicolumn{1}{c}{GFMC} & \multicolumn{1}{c}{Empirical}
          & \multicolumn{1}{c}{Experiment} 
          & \multicolumn{1}{c}{$r_p$} \\
\hline
$0^+$     &  -56.3(1)  &             &  -56.50      &  2.40 \\
$2^+$     &  + 3.2(2)  &             &  + 3.03(1)   &  2.45(1) \\
$4^+$     &  +11.2(3)  &             &  +11.35(15)  &  2.48(2) \\
$2^+;0$   &  +16.8(2)  &  +16.746(3) &  +16.626(3)  &  2.28 \\
$2^+;1$   &  +16.8(2)  &  +16.802(3) &  +16.922(3)  &  2.33 \\
$1^+;1$   &  +17.5(2)  &  +17.66(1)  &  +17.640(1)  &  2.39 \\
$1^+;0$   &  +18.0(2)  &  +18.13(1)  &  +18.150(4)  &  2.36 \\
$3^+;1$   &  +19.4(2)  &  +19.10(3)  &  +19.07(3)   &  2.31 \\
$3^+;0$   &  +19.9(2)  &  +19.21(2)  &  +19.235(10) &  2.35
\label{tab:energy}
\end{tabular}
\end{ruledtabular}
\end{table}

Barker evaluated the Coulomb contribution to the IM matrix element
$H^C_{01}$ in all three cases and found it to have the correct negative
sign, but only about half the required magnitude~\cite{Barker66}.
A variational Monte Carlo (VMC) evaluation of the mixing using the microscopic
Argonne v$_{18}$ (AV18) nucleon-nucleon interaction, which
has additional electromagnetic terms and strong charge-independence breaking, 
found significant additional contributions to $H_{01}$~\cite{WPCP00}.
In this paper we carry out more accurate Green's function Monte Carlo
(GFMC) evaluations of these terms, and consider extensions of the
charge-independence breaking of the original AV18 model.
We also evaluate the mixing matrix element with the first $2^+$ state 
of $^8$Be, which is the final state for the beta-decay of both $^8$Li and
$^8$B and a testing ground for weak decay terms beyond the Standard Model.

\section {Hamiltonian}
\label{sec:ham}

Charge symmetry implies the invariance of a system under a rotational transformation
which reverses the sign of the third component of isospin for all its
components, e.g., in nuclei $p \rightarrow n$ and $n \rightarrow p$.
The classification of $N\!N$ forces according to their dependence on
isospin or charge has been given by Henley and Miller~\cite{HM79}.
The dominant $N\!N$ forces are class I or charge-independent (CI) forces,
which may depend on the total isospin of a pair, but not on the charges
of the individual nucleons.
Thus in a $T$=1 state, a CI force between $pp$, $np$, and $nn$ pairs
is identical, while the force for a $T$=0 $np$ pair can be different.
A class II force is charge-dependent (CD) but maintains charge symmetry,
so in $T$=1 states, a CD force for $pp$ and $nn$ pairs is identical, 
but different for $np$ pairs.
Both class III and class IV potentials violate charge independence
and charge symmetry, with a class III charge-symmetry-breaking (CSB) force
differentiating between $pp$ and $nn$ pairs, while a class IV force
can mix $T$=0 and 1 $np$ pairs.
The Coulomb force between two protons can be written as a linear
combination of class I, II, and III terms, while the interaction
between nucleon magnetic moments involves all four classes.
The relative magnitude of these $N\!N$ forces is in the order
class I $>$ II $>$ III $>$ IV~\cite{vKFG96}.

The Hamiltonian used in this work has the form
\begin{equation}
H = \sum_{i} K_i + {\sum_{i<j}} v_{ij} + \sum_{i<j<k} V_{ijk} \ ,
\end{equation}
where $K_i$ is the nonrelativistic kinetic energy and $v_{ij}$ and $V_{ijk}$
are respectively the Argonne $v_{18}$ (AV18) \cite{WSS95} and Illinois-7
(IL7) \cite{PPWC01,P08b} potentials.
The kinetic energy includes both CI and CSB contributions,
the latter coming from the neutron-proton mass difference:
\begin{eqnarray}
  K_{i} & = & K^{\rm CI}_{i} + K^{\rm CSB}_{i} \\
        & = & -\frac{\hbar^2}{4}
     (\frac{1}{m_{p}} + \frac{1}{m_{n}}) \nabla^{2}_{i}
    -\frac{\hbar^2}{4} (\frac{1}{m_{p}} - \frac{1}{m_{n}})\tau_{iz}
     \nabla^{2}_{i} \nonumber \ ,
\end{eqnarray}
where $\tau_{iz}$ is the third component of isospin for nucleon $i$.
The AV18 potential has the structure:
\begin{equation}
  v_{ij} = v_{\gamma}(r_{ij})
         + \sum_{p=1,18} [ v^p_{\pi}(r_{ij}) + v^p_{I}(r_{ij})
         + v^p_{S}(r_{ij}) ] O^p_{ij} \ .
\end{equation}
Here $v_{\gamma}$ is a complete electromagnetic interaction, including
Coulomb, magnetic moment, vacuum polarization, and other terms.
The nuclear part of the potential has long-range one-pion-exchange (OPE)
$v_{\pi}$, and intermediate $v_{I}$ and short-range $v_{S}$
phenomenological parts.
The operators $O^p_{ij}$ include fourteen CI terms:
\begin{eqnarray}
  O^{p=1,14}_{ij} & = & [1, \sigma_{i}\cdot\sigma_{j}, S_{ij},
    {\bf L\cdot S},{\bf L}^{2},{\bf L}^{2}(\sigma_{i}\cdot\sigma_{j}),
    ({\bf L\cdot S})^{2}] \nonumber \\
                  & \otimes & [1,\tau_{i}\cdot\tau_{j}] \ ,
\end{eqnarray}
plus three CD terms and one class III CSB term:
\begin{equation}
O^{p=15,18}_{ij} = [1, \sigma_{i}\cdot\sigma_{j},
S_{ij}]\otimes T_{ij} \ , \ (\tau_i+\tau_j)_z \ .
\end{equation}
Here $\sigma_{i}$ is the Pauli spin operator for nucleon $i$,
$S_{ij} = 3(\sigma_{i}\cdot {\hat r}_{ij})(\sigma_{j}\cdot {\hat r}_{ij}) - \sigma_{i}\cdot\sigma_{j}$
is the tensor operator, ${\bf S} = (\sigma_{i} + \sigma_{j})/2$ is the total
pair spin, ${\bf L}$ is the pair orbital momentum operator,
and $T_{ij} = 3\tau_{iz}\tau_{jz} - \tau_{i}\cdot\tau_{j}$ is the
isotensor operator.

The long-ranged OPE yields a significant CD term arising 
from the difference between neutral and charged-pion masses.
The intermediate and short-range contributions to the force are constrained by
the differences between the considerable amount of $pp$ and $np$ scattering 
data in the $^1S_0$ channel.
Additional charge dependence, such as that arising from a spin-orbit term, 
might be expected.
Extracting such a term would require an independent analysis of $np$ data in
$^3P_J$ channels, which has not yet been made available.

The CSB term was determined by a slight alteration of the $^1S_0(pp)$
potential to fit the only available piece of $nn$ scattering data, the 
singlet scattering length $^1a_{nn}$.
When AV18 was constructed, the best data for $^1a_{nn}$ came
from $\pi^- d \rightarrow nn\gamma$ experiments, with a deduced
value of $-18.5(4)$ fm~\cite{TG87}; subsequent experiments and
analyses have not changed this significantly, with a current best
value of $-18.6(4)$ fm~\cite{MS01}.
The difference between the strong $pp$ and $nn$ scattering lengths in AV18,
i.e., after removal of the electromagnetic contributions, is 1.65 fm,
so the experimental error bar suggests an uncertainty in the strong
CSB term of order 25\%.

A major source for the nuclear CSB term is expected to be mixed 
$\pi$-$\eta$-$\eta^\prime$ and $\rho$-$\omega$ meson exchanges, with
the latter heavy-meson term dominant~\cite{HM79}.
Consequently only the short-range $v_{S}$
part of AV18 was altered, with the added assumption of spin-independence.
Again, one might well expect there to be additional CSB terms,
of spin-spin, tensor, and spin-orbit character, but 
additional $nn$ scattering data would be required to identify them
empirically.

A more complete model for $\rho$-$\omega$ exchange has been discussed
by Friar and Gibson \cite{FG78} (hereafter FG) and we will use a slightly
simplified version as an alternative to the single CSB term from AV18 above.
FG describe their model as a supplement to earlier work by McNamee, Scadron,
and Coon~\cite{MSC75}.
We wish to have a local potential for our many-body calculations, so we
neglect terms quadratic in momentum and reduce Eq.(9) in FG to the 
following form:
\begin{eqnarray}
\label{eq:rho-omega}
  v_{\rho\-\omega} = \big\{ v & + & \frac{1}{4M^2} (\mu_V + \mu_S) \nabla^2 v \nonumber \\*
        & + & \frac{1}{4M^2} \frac{2 \mu_V \mu_S}{3} \nabla^2 v ~ \sigma_{i}\cdot\sigma_{j} \nonumber \\*
        & - & \frac{1}{4M^2} \frac{\mu_V \mu_S}{3} [\nabla^2 v 
          - \frac{3v^\prime}{r}] S_{ij} \\*
        & + & \frac{1}{4M^2} ~ 4(\mu_V + \mu_S) ~ \frac{v^\prime}{r} ~ {\bf L\cdot S} \big\} ~ (\tau_i+\tau_j)_z \nonumber \\*
        & + & \frac{1}{4M^2} ~ 4(\mu_V - \mu_S) ~ \frac{v^\prime}{r} ~ {\bf L\cdot A} ~ (\tau_i-\tau_j)_z \ . \nonumber 
\end{eqnarray}
Here $\mu_S = 1 + \kappa_S$ and $\mu_V = 1 + \kappa_V$, where $\kappa$ is
the ratio of tensor to vector couplings of the isoscalar $\omega$ and
isovector $\rho$ mesons and $M = 938.9$ MeV is the average nucleon mass.
The first four lines are class III CSB terms, while the last line is 
an antisymmetric spin-orbit term with ${\bf A} = (\sigma_{i} - \sigma_{j})/2$
that is a class IV CSB contribution.

We emphasize here that our object is to explore the consequences of having
a more complete operator structure for CSB than AV18, one that acts differently
in different partial waves.
However, we would like to use a form consistent with AV18 for these new terms,
so instead of using an explicit heavy-meson exchange, we adopt the same 
short-range behavior for $v$, i.e., a modified Woods-Saxon potential with 
zero slope at the origin:
\begin{equation}
  v(r) = C (1 + F r)W(r) \ ,
\end{equation}
where $C$ is an overall strength factor adjusted to reproduce $^1a_{nn}$, and
\begin{eqnarray}
  W(r) & = & 1/\{1 + {\rm exp}[(r-R)/a]\} \ , \\
  F    & = & \frac{{\rm exp}(-R/a)}{a[1+{\rm exp}(-R/a)]} \ .
\end{eqnarray}
We set $R$ = 0.5 fm and $a$ = 0.2 fm as in the original AV18.
This choice of radial form for $v$ has the useful feature that both
$(v^\prime/r)$ and $\nabla^2 v$ in Eq.(\ref{eq:rho-omega}) remain finite
and well-behaved at the origin.

The form above gives a specific estimate for the relative strengths of 
the central, spin-spin, tensor, and spin-orbit CSB terms, once values
for $\kappa_S$ and $\kappa_V$ are specified.
We will consider two variations of this model, with ``small" and ``large"
values of the constants, designated AV18(s) and AV18(L),
as suggested by FG and Williams, Thomas, and Miller~\cite{WTM87}
(hereafter WTM), respectively.
The values for $C$, $\kappa_S$, and $\kappa_V$ are given in 
Table~\ref{tab:kappa} along with the $nn$ scattering lengths.
The $pp$ and $np$ scattering lengths of AV18 are unchanged with these
model variations.

\begin{table}[t]
\caption{Values of constants for different $v_{\rho\-\omega}$ models
and the resultant $nn$ scattering length.}
\begin{ruledtabular}
\begin{tabular}{ l d d d d }
Model & \multicolumn{1}{r}{$C$ (MeV)} & \multicolumn{1}{c}{$\kappa_V$}
      & \multicolumn{1}{c}{$\kappa_S$} & \multicolumn{1}{c}{$^1a_{nn}$} \\
\hline
AV18    & 0.98025 & 0.  & 0.   & -18.487 \\
AV18(s) & 1.11160 & 3.7 &-0.12 & -18.488 \\
AV18(L) & 1.50875 & 6.1 & 0.14 & -18.494
\label{tab:kappa}
\end{tabular}
\end{ruledtabular}
\end{table}

\begin{figure}[!t]
\epsfig{file=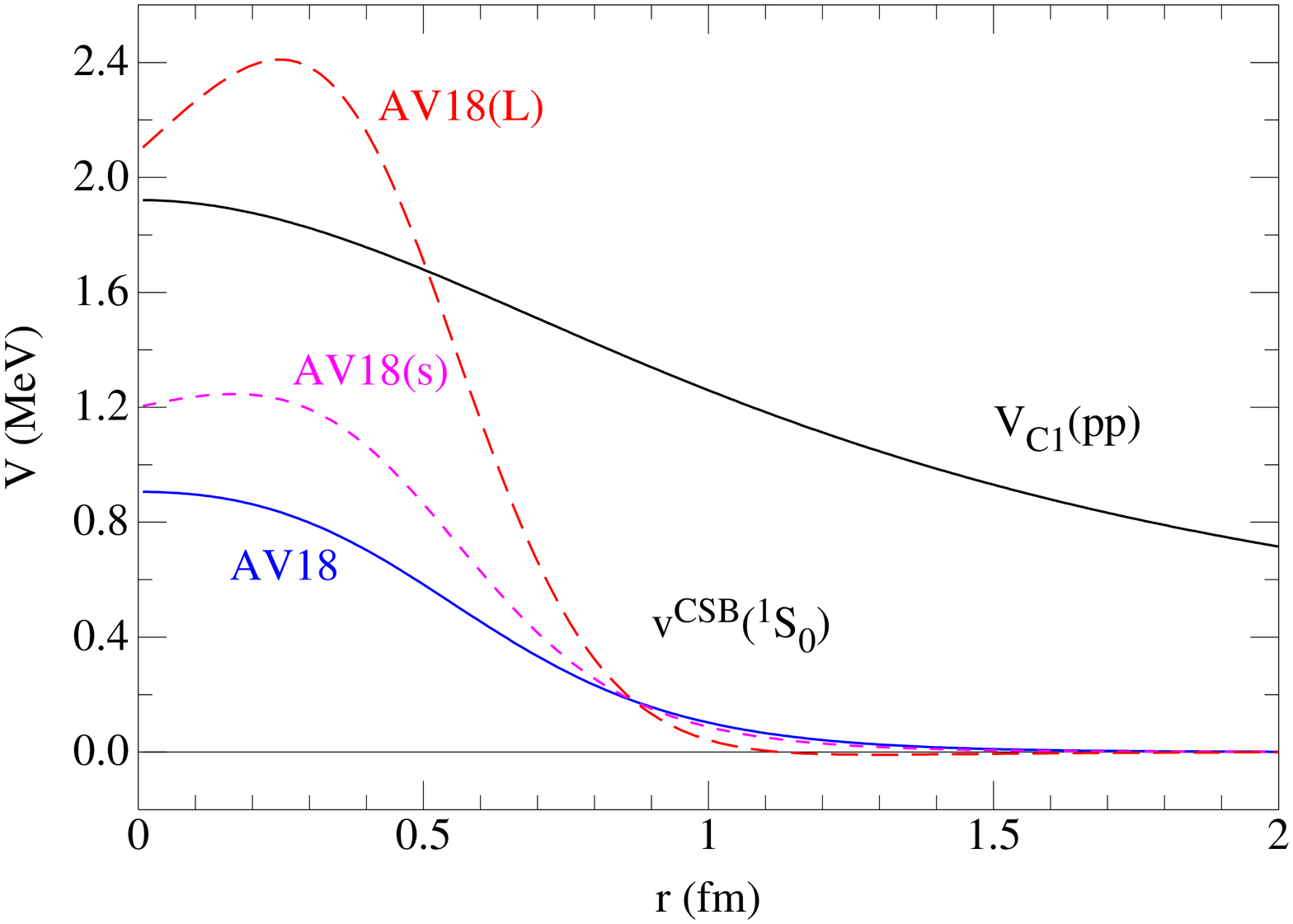,angle=270,width=8.0cm}
\caption{(Color online) Radial shapes of $v^{\rm CSB}$ in the $^1$S$_0$
channel for the three different models AV18 (solid blue line), 
AV18(s) (short-dash magenta line), and AV18(L) (long-dash red line),
compared to the proton-proton Coulomb interaction with form factor
(solid black line).}
\label{fig:1s0}
\end{figure}

The $v^{\rm CSB}$ in the $^1$S$_0$ channel for the three different models 
are shown in Fig.~\ref{fig:1s0} where they are compared with the static 
Coulomb potential $V_{C1}(pp)$ with the form factor used in AV18.

For the class IV CSB forces, we use the work of WTM
who studied CSB in neutron-proton elastic scattering, where these
forces can produce a difference in $n$ and $p$ analyzing powers.
The parameters used by WTM lead to values of CSB analyzing power differences
that are consistent with the TRIUMF~\cite{TRIUMF} and  IUCF~\cite{IUCF} measurements.
WTM identify one-photon-, one-pion-, and one-rho-exchange (ORE)
contributions to class IV CSB terms, in addition to the rho-omega
mixing term above.
The one-photon-exchange term acts only between $np$ pairs and can be written as:
\begin{equation}
\label{eq:gamma}
  v^{\rm IV}_\gamma = \alpha ~ \frac{\mu_n}{2M^2} ~ \frac{F_{ls}(r)}{r^3} ~ {\bf L\cdot A} ~ (\tau_i - \tau_j)_z \ ,
\end{equation}
where $\mu_n = -1.91$ n.m.\ is the neutron magnetic moment
and $F_{ls}(r)$ is a form factor for the finite size of the nucleon.
This is just the antisymmetric spin-orbit part of $V_{MM}(np)$ of AV18, 
Eq.(15) of \cite{WSS95}, with the form factor given in Eq.(10);
it is also equivalent to Eq.(3.3) of WTM.
The OPE and ORE terms are of the form:
\begin{eqnarray}
\label{eq:pi+rho}
  v^{\rm IV}_{\pi+\rho} & = & v_\rho(r) ~ {\bf L}\cdot{\bf A} ~ (\tau_i - \tau_j)_z \\
               & + & \big\{ w_\pi(r) + w_\rho(r) \big\} 
    {\bf L}\cdot(\sigma_i\times\sigma_j) ~ (\tau_i\times\tau_j)_z \nonumber \ .
\end{eqnarray}
The OPE radial function is given by:
\begin{equation}
  w_\pi(r) = \frac{g^2_\pi}{4\pi} ~ \frac{m^2_\pi}{2M^2} ~ 
             \frac{(m_n-m_p)}{2M} ~ m_{\pi} ~ Z_\pi(r) \ ,
\end{equation}
where $m_\pi = 139.6$ MeV is the charged pion mass and
\begin{equation}
  Z_x(r) = [ \frac{1}{\mu r} + \frac{1}{(\mu r)^2} ] 
               \frac{e^{-\mu r}}{\mu r} (1 - e^{-c_x r^2})^{3/2} \ ,
\end{equation}
with $\mu = m_x/\hbar c$.
This is equivalent to Eq.(3.10) of WTM, with a form factor chosen so that 
$Z_x(r)$ goes to a constant as $r \rightarrow 0$.
To be consistent with the OPE part of AV18, we take $c_\pi = 2.1$ fm$^{-2}$
and $g^2_\pi/4\pi = (2Mf/m_\pi)^2$ with $f^2 = 0.075$.
For rho-meson exchange there are both scalar and tensor terms:
\begin{eqnarray}
  v_\rho(r) & = & \frac{g_\rho^2}{4\pi} ~ \frac{m^2_{\rho}}{2M^2} ~
                  \frac{(m_n-m_p)}{2M} ~ m_\rho ~ Z_\rho(r) \ , \\
  w_\rho(r) & = & (1+\kappa_V)^2 ~ v_\rho(r) \ .
\end{eqnarray}
We use $m_\rho = 770$ MeV, the coupling constant $g_\rho^2/4\pi = 0.55$,
and the form factor cutoff $c_\rho = 2.44$ fm$^{-2}$.

\section {Quantum Monte Carlo method}
\label{sec:qmc}

We seek accurate solutions of the many-nucleon Schr\"{o}dinger equation
\begin{equation}
  H \Psi(J^\pi;T,T_z)= E \Psi(J^\pi;T,T_z) \ ,
\end{equation}
where $\Psi(J^\pi;T,T_z)$ is a nuclear wave function with specific spin-parity
$J^\pi$, isospin $T$, and charge state $T_z$.
We begin with a variational Monte Carlo (VMC) calculation, constructing 
a variational function $\Psi_V(J^\pi;T,T_z)$
from products of two- and three-body correlation operators acting on an
antisymmetric single-particle state of the appropriate quantum numbers.
The correlation operators are designed to reflect the influence of the
interactions at short distances, while appropriate boundary conditions
are imposed at long range~\cite{W91,PPCPW97}.
The $\Psi_V(J^\pi;T)$ has embedded variational parameters
that are adjusted to minimize the expectation value
\begin{equation}
 E_V = \frac{\langle \Psi_V | H | \Psi_V \rangle}
            {\langle \Psi_V   |   \Psi_V \rangle} \geq E_0 \ ,
\label{eq:expect}
\end{equation}
which is evaluated by Metropolis Monte Carlo integration~\cite{MR2T2}.
Here $E_0$ is the exact lowest eigenvalue of $H$ for the specified quantum 
numbers.
A good variational trial function has the form
\begin{equation}
   |\Psi_V\rangle =
      {\cal S} \prod_{i<j}^A
      \left[1 + U_{ij} + \sum_{k\neq i,j}^{A}\tilde{U}^{TNI}_{ijk} \right]
      |\Psi_J\rangle \ ,
\label{eq:psit}
\end{equation}
where the ${\cal S}$ is a symmetrization operator.
The Jastrow wave function $\Psi_J$ is fully antisymmetric and has the
$(J^\pi;T,T_z)$ quantum numbers of the state of interest, while $U_{ij}$
and $\tilde{U}^{TNI}_{ijk}$ are the two- and three-body correlation
operators.
Although we construct the $\Psi_V(J^\pi;T,T_z)$ to be an eigenstate of
the isospin $T$, we allow isobaric analog states with different $T_z$ to have
different wave functions, reflecting primarily the difference in
Coulomb contributions.

The GFMC method~\cite{C87,C88} improves on the VMC wave functions by acting
on $\Psi_V$ with the operator $\exp \left[-\left(H - E_0\right)\tau\right]$.
In practice, a simplified version $H^\prime$ of the Hamiltonian $H$
is used in the operator, which includes the isoscalar part of the
kinetic energy, a charge-independent eight-operator projection of AV18 called
AV8$^\prime$, a strength-adjusted version of the three-nucleon potential
IL7$^\prime$ (adjusted so that $\langle H^\prime \rangle \sim \langle H \rangle$),
and an isoscalar Coulomb term that integrates to the total charge of the
given nucleus~\cite{KNBSK99}.
The difference between $H$ and $H^\prime$ is calculated using perturbation
theory.
More detail can be found in Refs.~\cite{PPCPW97,WPCP00}.

The operator is applied in small slices of imaginary
time $\tau$ to produce a propagated wave function:
\begin{eqnarray}
  \Psi(\tau) = e^{-({H^\prime}-E_0)\tau} \Psi_V
             = \left[e^{-({H^\prime}-E_0)\triangle\tau}\right]^{n} \Psi_V \ .
\end{eqnarray}
Obviously $\Psi(\tau=0) = \Psi_V$ and $\Psi(\tau \rightarrow \infty) = \Psi_0$.
The algorithm for propagation produces samples of the wave function
$\Psi(\tau)$ but does not provide gradient information.
Quantities of interest are evaluated in terms of a ``mixed''
expectation value between $\Psi_V$ and $\Psi(\tau)$:
\begin{eqnarray}
  \langle O(\tau) \rangle_M & = & \frac{\langle \Psi(\tau) | O |\Psi_V
  \rangle}{\langle \Psi(\tau) | \Psi_V\rangle},
\label{eq:expectation}
\end{eqnarray}
where the operator $O$ acts on the trial function $\Psi_V$.
The desired expectation values, of course, have $\Psi(\tau)$ on both
sides; by writing $\Psi(\tau) = \Psi_V + \delta\Psi(\tau)$  and neglecting
terms of order $[\delta\Psi(\tau)]^2$, we obtain the approximate expression
\begin{eqnarray}
  \langle O (\tau)\rangle &=&
  \frac{\langle\Psi(\tau)| O |\Psi(\tau)\rangle}
  {\langle\Psi(\tau)|\Psi(\tau)\rangle}  \nonumber \\
  &\approx& \langle O (\tau)\rangle_M
    + [\langle O (\tau)\rangle_M - \langle O \rangle_V] ~,
\label{eq:pc_gfmc}
\end{eqnarray}
where $\langle O \rangle_{\rm V}$ is the variational expectation value.

For off-diagonal matrix elements relevant to this work the 
generalized mixed estimate is given by the expression
\begin{eqnarray}
&& \frac{\langle\Psi^f(\tau)| O |\Psi^i(\tau)\rangle}{\sqrt{\langle \Psi^f(\tau) | \Psi^f(\tau)\rangle}
\sqrt{\langle \Psi^i(\tau) |\Psi^i(\tau)\rangle}} \nonumber \\
&\approx&
  \langle O(\tau) \rangle_{M_i}
+ \langle O(\tau) \rangle_{M_f}-\langle O \rangle_V \ ,
\label{eq:extrap}
\end{eqnarray}
where
\begin{eqnarray}
\langle O(\tau) \rangle_{M_f}
& = & \frac{\langle \Psi^f(\tau) | O |\Psi^i_V\rangle}
           {\langle \Psi^f(\tau)|\Psi^f_V\rangle}
      \sqrt{\frac{\langle \Psi^f_V|\Psi^f_V\rangle}
           {\langle \Psi^i_V | \Psi^i_{V}\rangle}} \ ,
\label{eq:mixed_f}
\end{eqnarray}
and $\langle O(\tau) \rangle_{M_i}$ is defined similarly.
For more details see Eqs.~(19-24) and the accompanying discussions in 
Ref.~\cite{PPW07}.

\begin{table}[t]
\caption{Contributions to the isovector energy difference 
$E(^3{\rm He})-E(^3{\rm H})$ in keV from different interaction models; 
Monte Carlo statistical errors are shown in parentheses.}
\label{tab:A3}
\begin{ruledtabular}
\begin{tabular}{l r r r r}
             &  \multicolumn{1}{c}{AV18}
             &  \multicolumn{1}{c}{AV18(s)}
             &  \multicolumn{1}{c}{AV18(L)} \\
\hline
                    &        &        & \vspace{-8pt}\\
$K^{\rm CSB}$       &  14(0) &  14(0) &  14(0)  \\
$V_{C1}(pp)$        & 642(1) & 642(1) & 642(1)  \\
$V_{C+}$            &   9(0) &   9(0) &   9(0)  \\
$V_{MM}$            &  17(0) &  17(0) &  17(0)  \\
$v^{\rm CSB}$       &  65(0) &  71(0) &  79(0)  \\
$\delta H^{\rm CI}$ &   8(1) &   8(1) &   8(1)  \\
\hline
                    &        &        & \vspace{-8pt}\\
$H^{\rm CSB}$       & 755(1) & 761(1) & 769(1)  \\
\hline
                    &        &        & \vspace{-8pt}\\
Experiment          &        &        & 764~~~~
\end{tabular}
\end{ruledtabular}
\end{table}

\section {Results}
\label{sec:res}

\subsection {Energies of Ground and Excited States in $^8$Be}

The GFMC energy for the ground state and the excitation energies
for the first eight positive-parity excited states of $^8$Be are
given in Table~\ref{tab:energy}; these have been calculated for 
pure isospin states of either $T$=0 or $T$=1.
The experimentally observed energies and the empirical energies
for the unmixed states (derived as discussed in Sec.~\ref{sec:intro})
are also shown, along with the GFMC rms point proton radii.
As discussed in~\cite{Datar13}, the physically wide $2^+$ and $4^+$
states present a challenge for GFMC calculations because they
tend to break up into separate $\alpha$ particles as the propagation
in imaginary time $\tau$ proceeds.
The energy drifts lower and the radii increase with $\tau$, so care
is necessary to extract these quantities from the calculations.
However, no such problem occurs for the physically much narrower $2^+$, 
$1^+$ and $3^+$ doublets; their GFMC energies and radii are quite
stable as $\tau$ increases.

The AV18+IL7 Hamiltonian reproduces both the 2$\alpha$-like $0^+$, $2^+$,
$4^+$ rotational band and the mixed $2^+$ and $1^+$ doublets exceptionally
well
Only the $3^+$ doublet is about 0.5 MeV too high in excitation energy, 
and with perhaps too big an energy difference between the states.
The radii of the mixed doublets are all slightly smaller than the 
2$\alpha$-like states.
The energies and other properties of the isobaric analog states in 
$^8$Li and $^8$B are also in good agreement with experiment
for this Hamiltonian~\cite{PPSW13}.

\subsection {Isovector Energy Differences of Mirror Nuclei}

We next examine the effect of our different interaction models on the
isovector energy differences in $A$=3, $A$=7,$T$=$\frac{1}{2}$, and $A$=8,$T$=1
mirror nuclei.
The energy difference $E(^3{\rm He})-E(^3{\rm H})$ for two correlated
GFMC propagations~\cite{BPW11} is given in Table~\ref{tab:A3}.
The starting variational wave functions were separately optimized for
the two different charge states with the proper experimental charge radii.
As stated above, the propagation is made with AV8$^\prime$+IL7$^\prime$
plus an isoscalar Coulomb term that integrates to the proper total charge
for each nucleus, and the difference with AV18+IL7 or the variants of 
AV18 is evaluated using perturbation theory.

\begin{table}[t]
\caption{Contributions to the isovector energy difference 
$E(^7{\rm Be})-E(^7{\rm Li})$ in keV from different interaction models; 
Monte Carlo statistical errors are shown in parentheses.}
\label{tab:A7}
\begin{ruledtabular}
\begin{tabular}{l r r r}
             &  \multicolumn{1}{c}{AV18}
             &  \multicolumn{1}{c}{AV18(s)}
             &  \multicolumn{1}{c}{AV18(L)} \\
\hline
                    &           &           & \vspace{-8pt}\\
$K^{\rm CSB}$       &    23(0)~~&    23(0)~~&    23(0)~~\\
$V_{C1}(pp)$        &  1442(2)~~&  1442(2)~~&  1442(2)~~\\
$V_{C+}$            &    18(0)~~&    18(0)~~&    18(0)~~\\
$V_{MM}$            &    18(0)~~&    18(0)~~&    18(0)~~\\
$v^{\rm CSB}$       &    83(1)~~&    90(1)~~&   105(1)~~\\
$\delta H^{\rm CI}$ &    27(10) &    27(10) &    27(10) \\
\hline
                    &           &           & \vspace{-8pt}\\
$H^{\rm CSB}$       &  1611(10) &  1618(10) &  1633(10) \\
\hline
                    &           &           & \vspace{-8pt}\\
Experiment          &           &           &  1645~~~~~
\end{tabular}
\end{ruledtabular}
\end{table}

The different contributions include 1) the kinetic energy $K^{\rm CSB}$,
2) the static Coulomb term between two protons (with finite-range form
factor) $V_{C1}(pp)$, 3) all additional charge contributions to the
electromagnetic interaction $V_{C+}$ (like Darwin-Foldy and vacuum
polarization), 4) the magnetic moment term $V_{MM}$, and 5) the strong
class III CSB term $v^{\rm CSB}$, which is the single term from AV18
or the sum of the first four rows of Eq.(\ref{eq:rho-omega}) for AV18(s)
and AV18(L).
The net change in the energy arising from the CI part of the Hamiltonian 
$\delta H^{\rm CI}$ is an additional second-order perturbation correction
due to differences in the two GFMC propagations.
This term is small for the $A$=3 case, although
the changes in separate kinetic and potential parts are much larger.

(We note that the original $V_{C1}(pp)$ from Eq.(4) of Ref.~\cite{WSS95}
uses an $\alpha^\prime$ that has a small relativistic energy-dependence;
we drop this from the many-body calculations and just use $\alpha$, which
is what we mean by the term ``static" Coulomb.
However, we add a momentum-dependent orbit-orbit term to the $V_{C+}$,
which is typically $\sim$1\% of the static term, to approximate this term.) 

The dominant contribution to the isovector energy difference is of course 
the static Coulomb interaction between 
protons, which is in agreement with model-independent estimates based on the 
experimental form factors~\cite{BCS78}.
Comparing the three different models for $v^{\rm CSB}$, the smallest is AV18, 
with AV18(s) being about 10\% larger and AV18(L) about 20\% larger, as
might be expected from the larger size of $v^{CSB}$ in the $^1$S$_0$
channel shown in Fig.~\ref{fig:1s0}.
All the models give reasonable agreement with the experimental difference 
of 764 keV.

The isovector energy differences for the $A$=7,$T$=$\frac{1}{2}$ mirror nuclei 
are shown in Table~\ref{tab:A7}.
Again we show the difference of two correlated GFMC propagations~\cite{BPW11}
that have been started from separately optimized 
variational trial functions.
The change in the CI part of the Hamiltonian $\delta H^{\rm CI}$
is larger than for $A$=3 and has a larger error bar, which now dominates
the total error in $H^{\rm CSB}$.
The net GFMC results are a little smaller than the experimental
energy difference of 1645 keV.
The $v^{\rm CSB}$ again increases about 20\% going from AV18 to AV18(s) 
to AV18(L) as in the $A$=3 case, because the CSB force is being probed 
primarily in $S$=0, $T$=1 pairs embedded in the $p$-shell~\cite{W06}.

Finally, the $A$=8,$T$=1 isovector energy difference 
$[E(^8{\rm B})-E(^8{\rm Li})]/2$ is shown in Table~\ref{tab:A8}.
The static Coulomb contribution is similar to the $A$=7 case, while
the magnetic moment contribution almost vanishes.
Notably, the variation between the AV18, AV18(s), and AV18(L) models is
different, probably because there are now equal numbers of $S$=0 and 1,
$T$=1 pairs embedded in the $p$-shell~\cite{W06} and the spin-dependence
of Eq.(\ref{eq:rho-omega}) comes into play.
The change in the CI Hamiltonian $\delta H^{\rm CI}$ is similar to that
in $A$=7 and the total $H^{\rm CSB}$ is somewhat
over-predicted compared to the experimental value of 1770 keV.

\begin{table}[t]
\caption{Contributions to the isovector energy difference 
$[E(^8{\rm B})-E(^8{\rm Li})]/2$ in keV from different interaction models; 
Monte Carlo statistical errors are shown in parentheses.}
\label{tab:A8}
\begin{ruledtabular}
\begin{tabular}{l r r r}
             &  \multicolumn{1}{c}{AV18}
             &  \multicolumn{1}{c}{AV18(s)}
             &  \multicolumn{1}{c}{AV18(L)} \\
\hline
                    &           &           & \vspace{-8pt}\\
$K^{\rm CSB}$       &    25(0)~~&    25(0)~~&    25(0)~~\\
$V_{C1}(pp)$        &  1652(3)~~&  1652(3)~~&  1652(3)~~\\
$V_{C+}$            &    17(0)~~&    17(0)~~&    17(0)~~\\
$V_{MM}$            &     1(0)~~&     1(0)~~&     1(0)~~\\
$v^{\rm CSB}$       &    77(1)~~&    75(2)~~&    84(3)~~\\
$\delta H^{\rm CI}$ &    33(11) &    33(11) &    33(11) \\
\hline
                    &           &           & \vspace{-8pt}\\
$H^{\rm CSB}$       &  1813(11) &  1811(11) &  1820(11) \\
\hline
                    &           &           & \vspace{-8pt}\\
Experiment          &           &           &  1770~~~~~
\end{tabular}
\end{ruledtabular}
\end{table}

In all three pairs of mirror nuclei, the static Coulomb 
interaction between protons is the dominant source of the energy difference, 
providing about 85-90\% of the total, increasing as $A$ increases.
The kinetic and remaining electromagnetic terms provide another few percent,
leaving the remaining amount due to strong CSB terms.
However, these terms are of shorter range than Coulomb, and their
total contributions do not grow as rapidly with $A$, so they
become relatively less important in larger nuclei.

\subsection {Isospin-Mixing Matrix Elements in $^8$Be}

The GFMC evaluation for the IM matrix element $H_{01}$
between the $2^+$ states at 16.6--16.9 MeV excitation in $^8$Be
is given in Table~\ref{tab:A8-2}.
The first five lines give the contributions for the same terms as in the energy 
differences for mirror nuclei of Tables~\ref{tab:A3}--\ref{tab:A8}.
In addition there are rows for the additional class IV CSB terms:
$v^{\rm IV}_\gamma$ of Eq.(\ref{eq:gamma}), 
$v^{\rm IV}_{\pi+\rho}$ of Eq.(\ref{eq:pi+rho}),
and the antisymmetric spin-orbit term $v^{\rm IV}_{\rho\omega}$
that is the last line of Eq.(\ref{eq:rho-omega}).

We note that our Coulomb term of $-89$ keV is about 30\% larger than Barker's 
original estimate of $-67$ keV~\cite{Barker66}.  The additional
electromagnetic and kinetic terms that we include give $-15$ keV,
while the strong CSB terms add another $-24$ to $-28$ keV to the total.
Thus, the strong CSB terms are relatively more important here
than in the isovector energy differences between mirror nuclei, making
this system one of the best for constraining such forces.
The variation between AV18 and the alternative models AV18(s) and AV18(L) 
is proportionately the same as in the energy differences.
Our total of $-127$ to $-131$ keV is about 90\% of the empirical matrix 
element of $-145$ keV.

\begin{table}[t]
\caption{Contributions to the isospin-mixing matrix element between 
$^8$Be($2^+$) states at 16.6--16.9 MeV excitation for different interaction 
models in keV; Monte Carlo statistical errors are shown in parentheses.}
\label{tab:A8-2}
\begin{ruledtabular}
\begin{tabular}{l d d d}
             &  \multicolumn{1}{c}{AV18}
             &  \multicolumn{1}{c}{AV18(s)}
             &  \multicolumn{1}{c}{AV18(L)} \\
\hline
                           &  \multicolumn{3}{c}{} \vspace{-8pt}\\
$K^{\rm CSB}$              &  -3.6(1) &  -3.6(1) &  -3.6(1) \\
$V_{C1}(pp)$               & -89.3(11)& -89.3(11)& -89.3(11)\\
$V_{C+}$                   &   0.0(0) &   0.0(0) &   0.0(0) \\
$V_{MM}$                   & -10.2(2) & -10.2(2) & -10.2(2) \\
$v^{\rm CSB}$              & -23.4(4) & -24.7(6) & -25.7(10)\\
$v^{\rm IV}_{\gamma}$      &  -0.8(1) &  -0.8(1) &  -0.8(1) \\
$v^{\rm IV}_{\pi+\rho}$    &          &  -0.8(1) &  -0.8(1) \\
$v^{\rm IV}_{\rho\omega}$  &          &  -0.3(1) &  -0.8(3) \\
\hline
                           &  \multicolumn{3}{c}{} \vspace{-8pt}\\
$H_{01}$                   &-127.(2)  &-130.(2)  &-131.(2)  \\
\hline
                           &  \multicolumn{3}{c}{} \vspace{-8pt}\\
Experiment                 &          &          &-145.(3)
\end{tabular}
\end{ruledtabular}
\end{table}

\begin{table}[b]
\caption{Contributions to the isospin-mixing matrix element between 
$^8$Be($1^+$) states at 17.6--18.1 MeV excitation for different interaction 
models in keV; Monte Carlo statistical errors are shown in parentheses.}
\label{tab:A8-1}
\begin{ruledtabular}
\begin{tabular}{l d d d}
             &  \multicolumn{1}{c}{AV18}
             &  \multicolumn{1}{c}{AV18(s)}
             &  \multicolumn{1}{c}{AV18(L)} \\
\hline
                           &  \multicolumn{3}{c}{} \vspace{-8pt}\\
$K^{\rm CSB}$              &  -2.8(1) &  -2.8(1) &  -2.8(1) \\
$V_{C1}(pp)$               & -73.4(11)& -73.4(11)& -73.4(11)\\
$V_{C+}$                   &   0.0(0) &   0.0(0) &   0.0(0) \\
$V_{MM}$                   &  -1.1(1) &  -1.1(1) &  -1.1(1) \\
$v^{\rm CSB}$              & -18.5(4) & -18.7(6) & -19.9(10)\\
$v^{\rm IV}_{\gamma}$      &   2.1(1) &   2.1(1) &   2.1(1) \\
$v^{\rm IV}_{\pi+\rho}$    &          &   0.2(0) &   0.2(0) \\
$v^{\rm IV}_{\rho\omega}$  &          &   0.5(1) &   1.1(2) \\
\hline
                           &  \multicolumn{3}{c}{} \vspace{-8pt}\\
$H_{01}$                   & -94.(1)  & -93.(2)  & -94.(2)  \\
\hline
                           &  \multicolumn{3}{c}{} \vspace{-8pt}\\
Experiment                 &          &          &-103.(14)
\end{tabular}
\end{ruledtabular}
\end{table}

The GFMC evaluation for the IM matrix element $H_{01}$
between the $1^+$ states at 17.6--18.2 MeV excitation in $^8$Be
is given in Table~\ref{tab:A8-1}.
In this case, our $pp$ Coulomb term of $-73$ keV is 35\%
larger than Barker's estimate of $-54$ keV.
The strong class III CSB term is a little bit smaller than in the $2^+$ case. 
The magnetic moment term almost vanishes, and the class IV CSB terms
reduce the magnitude of the mixing matrix element.
This change of sign of the ${\bf L}\cdot{\bf A}$ terms relative to the
$2^+$ (and $3^+$) doublets is probably due to the significant admixture of 
$^1$P$_1$[431] symmetry components in the $T$=1 state.
In the end, the three interaction models again give a very
narrow spread of $-93$ to $-94$ keV, or about 90\% of the 
empirical value of $-103$ keV.

The GFMC evaluation for the IM matrix element $H_{01}$
between the $3^+$ states at 19.0--19.2 MeV excitation in $^8$Be
is given in Table~\ref{tab:A8-3}.
The Coulomb term at $-75$ keV is comparable to the previous cases, 
but now more than double Barker's estimate of $-32$ keV.
The strong class III CSB terms are similar to the previous cases.
The regular magnetic moment contribution is like that in the $2^+$ doublet,
but the class IV magnetic moment term is larger.
Overall, there is again very little spread between our models, at $-111$ to
$-115$ keV, but these are now much larger than the poorly determined
value of $-59$ keV for the empirical matrix element.
Further, the maximum possible IM matrix element is one half the spacing 
between the two physical states, which in this case is 165(32) keV.
The experimental energy difference would have to be about two standard 
deviations greater than given in the compilation~\cite{exp8-10} to admit 
an IM matrix element as large as that predicted by our Hamiltonian.
The GFMC calculations do hint at a bigger energy difference between
the isospin-pure states as shown in Table~\ref{tab:energy}.

\begin{table}[t]
\caption{Contributions to the isospin-mixing matrix element between 
$^8$Be($3^+$) states at 19.0--19.2 MeV excitation for different interaction 
models in keV; Monte Carlo statistical errors are shown in parentheses.}
\label{tab:A8-3}
\begin{ruledtabular}
\begin{tabular}{l d d d}
             &  \multicolumn{1}{c}{AV18}
             &  \multicolumn{1}{c}{AV18(s)}
             &  \multicolumn{1}{c}{AV18(L)} \\
\hline
                           &  \multicolumn{3}{c}{} \vspace{-8pt}\\
$K^{\rm CSB}$              &  -3.0(1) &  -3.0(1) &  -3.0(1) \\
$V_{C1}(pp)$               & -74.6(12)& -74.6(12)& -74.6(12)\\
$V_{C+}$                   &   0.0(0) &   0.0(0) &   0.0(0) \\
$V_{MM}$                   & -12.3(2) & -12.3(2) & -12.3(2) \\
$v^{\rm CSB}$              & -16.6(4) & -16.9(6) & -17.5(10)\\
$v^{\rm IV}_{\gamma}$      &  -4.5(1) &  -4.5(1) &  -4.5(1) \\
$v^{\rm IV}_{\pi+\rho}$    &          &  -0.3(0) &  -0.3(0) \\
$v^{\rm IV}_{\rho\omega}$  &          &  -1.3(1) &  -2.5(0) \\
\hline
                           &  \multicolumn{3}{c}{} \vspace{-8pt}\\
$H_{01}$                   &-111.(2)  &-112.(2)  &-115.(2)  \\
\hline
                           &  \multicolumn{3}{c}{} \vspace{-8pt}\\
Experiment                 &          &          & -59.(12)
\end{tabular}
\end{ruledtabular}
\end{table}

Finally we report the IM matrix element between the first $2^+$ $T$=0 state
in $^8$Be at 3.0 MeV excitation and the $2^+$ $T$=1 state at 16.8 MeV.
The former is the final state for $\beta$-decay from both $^8$Li and $^8$B 
ground states, while the latter is the isobaric analog of the initial states.
These transitions have been used to search for tensor components in
nuclear $\beta$-decay~\cite{bdecay13} and may be used in future experiments 
to search for second-class currents or other violations of the conserved 
vector-current hypothesis. 
The analysis of such experiments relies on the final state being a 
$T$=0 state with a negligible $T$=1 component.
Without giving the detailed breakdown, we get an IM matrix element of 
$-7(2)$ keV which, combined with the large separation in energy, implies
an amplitude admixture of the first $2^+$ $T$=1 state into the first
$2^+$ $T$=0 state of $5.0(1.5) \times 10^{-4}$.
This is sufficiently small to not interfere with the goal stated 
in Ref.~\cite{bdecay13} of improving the limit on tensor components by
an order of magnitude.

\vspace{0.25in}

\section {Conclusions}
\label{sec:conclusions}

In summary, we have reported GFMC results for IM matrix elements
in the $2^+$, $1^+$, and $3^+$ doublets at 16--19 MeV excitation
in the spectrum of $^8$Be.
We have made these calculations for the AV18+IL7 Hamiltonian and two
variants of AV18 with an expanded CSB operator structure, all constrained to
give the same $nn$ scattering length.
The AV18+IL7 model gives an excellent reproduction of the $^8$Be spectrum
and a good description of the energy differences in $A$=3, 7 and 8 mirror 
nuclei.
For the isospin-mixing matrix element, we add the class IV CSB terms 
that come from one-photon, one-pion, one-rho, and rho-omega mixing.
We have not considered possible three-nucleon CSB forces,
which have been estimated to contribute $\sim 5$ keV to the $^3$He--$^3$H
energy difference~\cite{FPvK05}.
We obtain about 90\% of the empirical IM matrix elements for the $2^+$
and $1^+$ doublets, but we overpredict the less well-measured $3^+$ doublet.
New experiments for these latter states would be useful.

Our main conclusion is that, while the static Coulomb interaction
between protons is the dominant contributor to CSB in nuclei,
the additional electromagnetic, kinetic, and strong CSB terms are
important, and the IM matrix elements in $^8$Be are a
particularly valuable place to look for their effect.

\begin{acknowledgments}
We thank I. Brida for valuable input.
The many-body calculations were performed on the parallel computers of the
Laboratory Computing Resource Center, Argonne National Laboratory.
The work of RBW and SCP is supported by the US DOE,
Office of Nuclear Physics, under contract No. DE-AC02-06CH11357
and by the NUCLEI SciDAC program,
that of SP by the US NSF under Grant No. PHY-1068305, and that of GAM
by US DOE under contract No. DE-FG02-97ER41014.
\end{acknowledgments}


\begin{thebibliography}{99}

\bibitem{exp8-10}
D. R. Tilley, J. H. Kelley, J. L. Godwin, D. J. Millener,
J. Purcell, C. G. Sheu, and H. R. Weller,
Nucl. Phys. {\bf A745}, 155 (2004).

\bibitem{W06}
R. B. Wiringa,
Phys. Rev. C {\bf 73}, 034317 (2006).

\bibitem{WPCP00}
R. B. Wiringa, S. C. Pieper, J. Carlson, and V. R. Pandharipande,
Phys. Rev. C {\bf 62}, 014001 (2000).

\bibitem{Barker66}
F. C. Barker,
Nucl. Phys. {\bf 83}, 418 (1966).

\bibitem{OG77}
M. A. Oothoudt and G. T. Garvey,
Nucl. Phys. {\bf A284}, 41 (1977).

\bibitem{HM79}
E. M. Henley and G. A. Miller, in {\em Mesons in Nuclei}, edited by M. Rho
and D. H. Wilkinson (North-Holland, Amsterdam, 1979), p. 405;
G. A. Miller, B. M. K. Nefkens and I. Slaus,
Phys. Rept. {\bf 194}, 1 (1990).

\bibitem{vKFG96}
U. van Kolck, J. L. Friar, and T. Goldman,
Phys. Lett. B {\bf 371}, 169 (1996).
 
\bibitem{WSS95}
R. B. Wiringa, V. G. J. Stoks, and R. Schiavilla,
Phys. Rev. C {\bf 51}, 38 (1995).

\bibitem{PPWC01}
S. C. Pieper, V. R. Pandharipande, R. B. Wiringa, and J. Carlson,
Phys. Rev. C {\bf 64}, 014001 (2001).

\bibitem{P08b}
S. C. Pieper,
AIP Conf. Proc. {\bf 1011}, 143 (2008).

\bibitem{TG87}
G. F. de T\'{e}ramond and B. Gabioud,
Phys. Rev. C {\bf 36}, 691 (1987).

\bibitem{MS01}
R. Machleidt and I. Slaus,
J. Phys. G {\bf 27}, R69 (2001).

\bibitem{FG78}
J. L. Friar and B. F. Gibson, 
Phys. Rev. C {\bf 17}, 1752 (1978).

\bibitem{MSC75}
P. C. McNamee, M. D. Scadron, and S. A. Coon,
Nucl. Phys. {\bf A249}, 483 (1975);
S. A. Coon, M. D. Scadron, and P. C. McNamee,
Nucl. Phys. {\bf A287}, 381 (1997).

\bibitem{WTM87}
A. G. Williams, A. W. Thomas, and G. A. Miller,
Phys. Rev. C {\bf 36}, 1956 (1987);
G. A. Miller, A. W. Thomas and A. G. Williams,
Phys. Rev. Lett. {\bf 56}, 2567 (1986).

\bibitem{TRIUMF}
R. Abegg, {\it et al.},
Phys. Rev. Lett.  {\bf 56}, 2571 (1986);
R. Abegg, {\it et al.},
Phys. Rev. D {\bf 39}, 2464 (1989);
R. Abegg, {\it et al.},
Phys. Rev. Lett.  {\bf 75}, 1711 (1995);
J. Zhao, {\it et al.},
Phys. Rev. C {\bf 57}, 2126 (1998).

\bibitem{IUCF}
L. D. Knutson, S. E. Vigdor, W. W. Jacobs, J. Sowinski, P. L. Jolivette, S. W. Wissink, C. Bloch, C. Whiddon, and R. C. Byrd,
Phys. Rev. Lett.  {\bf 66}, 1410 (1991);
S. E. Vigdor, W. W. Jacobs, L. D. Knutson, J. Sowinski, C. Bloch, P. L. Jolivette, S. W. Wissink, R. C. Byrd, and C. Whiddon,
Phys. Rev. C {\bf 46}, 410 (1992).

\bibitem{W91}
R. B. Wiringa,
Phys. Rev. C {\bf 43}, 1585 (1991).

\bibitem{PPCPW97}
B. S. Pudliner, V. R. Pandharipande, J. Carlson, S. C. Pieper, and
R. B. Wiringa,
Phys. Rev. C {\bf 56}, 1720 (1997).

\bibitem{MR2T2}
N. Metropolis, A. W. Rosenbluth, M. N. Rosenbluth, A. H. Teller, and E. Teller,
J. Chem. Phys. {\bf 21}, 1087 (1953).

\bibitem{C87}
J. Carlson,
Phys. Rev. C {\bf 36}, 2026 (1987).

\bibitem{C88}
J. Carlson,
Phys. Rev. C {\bf 38}, 1879 (1988).

\bibitem{KNBSK99}
G. P. Kamuntavi\v{c}ius, P. Navr\'{a}til, B. R. Barrett, G. Sapragonaite, and R. K. Kalinauskas,
Phys. Rev. C {\bf 60}, 044304 (1999).

\bibitem{PPW07}
M. Pervin, S. C. Pieper, and R. B. Wiringa,
Phys. Rev. C {\bf 76}, 064319 (2007).

\bibitem{BPW11}
I. Brida, S. C. Pieper, and R. B. Wiringa,
Phys. Rev. C {\bf 84}, 024319 (2011).

\bibitem{BCS78}
R. A. Brandenburg, S. A. Coon, and P. U. Sauer, 
Nucl. Phys. A {\bf 294}, 305 (1978).

\bibitem{Datar13}
V. M. Datar, {\it et al.},
Phys. Rev. Lett. {\bf 111}, 062502 (2013).

\bibitem{PPSW13}
S. Pastore, S. C. Pieper, R. Schiavilla, and R. B. Wiringa,
Phys. Rev. C {\bf 87}, 035503 (2013).

\bibitem{bdecay13}
G. Li, {\it et al.},
Phys. Rev. Lett. {\bf 110}, 092502 (2013).

\bibitem{FPvK05}
J. L. Friar, G. L. Payne, and U. van Kolck,
Phys. Rev. C {\bf 71}, 024003 (2005).

\end{thebibliography}
\end{document}